\documentclass{jetpl}
\twocolumn
\usepackage{bm}

\lat

\title{Physics of the Insulating Phase in the Dilute
  Two-Dimensional Electron Gas}

\rtitle{Physics of the Insulating Phase in the Dilute
  Two-Dimensional Electron Gas}

\sodtitle{Physics of the Insulating Phase in the Dilute
  Two-Dimensional Electron Gas}

\author{
Victor M.~Yakovenko$^*$\/\thanks{e-mail: yakovenk@physics.umd.edu},
Victor~A.~Khodel$^{+*}$}

\rauthor{YAKOVENKO, KHODEL}

\sodauthor{Yakovenko~V.~M., Khodel~V.~A.}

\address{$^*$Condensed Matter Theory Center, Department of Physics,
  University of Maryland, College Park,  Maryland 20742-4111, USA \\~\\
  $^+$Russian Research Centre Kurchatov Institute,  Moscow, 123182,
  Russia}

\dates{19 August 2003, published JETP Letters {\bf 78}, 398 (2003),}
{-- revised 17 January 2004, version 4 cond-mat/0308380}

\abstract{We propose to use the radio-frequency single-electron
  transistor as an extremely sensitive probe to detect the
  time-periodic ac signal generated by sliding electron lattice in the
  insulating state of the dilute two-dimensional electron gas.  We
  also propose to use the optically-pumped NMR technique to probe the
  electron spin structure of the insulating state.  We show that the
  electron effective mass and spin susceptibility are strongly
  enhanced by critical fluctuations of electron lattice in the
  vicinity of the metal-insulator transition, as observed in
  experiment.}

\PACS{73.20.Qt 
73.21.Fg 
73.20.Fz 
}

\begin{document}

\maketitle

\paragraph{Detecting electron lattice with the single-electron 
  transistor.}

The metal-insulator transition (MIT) in the two-dimensional electron
gas (2DEG) attracts considerable interest
\cite{Abrahams01,Kravchenko04}.  In this paper, we focus on physics of
the insulating phase.  The great majority of experiments are transport
measurements, and only few are thermodynamic.  Dultz and Jiang
\cite{Dultz} measured compressibility $\kappa$ of the 2DEG as a
function of carrier concentration $n$ and found that it tends to
vanish in the insulating phase, i.e.\ the phase is incompressible.
The experimental dependence $\kappa(n)$ was semi-quantitatively
reproduced within both the scenario of electron localization (E-LOC)
in disordered potential \cite{Shi-Xie} and the scenario of electron
lattice (E-LAT) formation \cite{Orozco}.  We use the term E-LAT to
denote any state with local periodic modulation of electron density.
The Wigner crystal (WC) and the charge-density wave (CDW) are the
limiting cases of E-LAT, where the modulation amplitude is comparable
to $n$ and is much less than $n$, respectively.  For simplicity, we
call the carriers electrons, even though they may actually be holes.

Ilani \emph{et al.} \cite{Yacobi00,Yacobi01} measured compressibility
locally, using the single-electron transistor (SET) as a microscopic
probe.  They found that $\mu(n)$ has a series of quasi-random jumps,
which become very strong in the insulating phase.  These jumps were
interpreted as single-electron charging events
\cite{Yacobi00,Yacobi01} within the E-LOC scenario.  Alternatively,
the jumps can be interpreted as a manifestation of E-LAT
\cite{Yacobi01}.  When the average carrier concentration $n$ is
changed by the back gate, the period $l$ of E-LAT must adjust, because
it is proportional to the average distance between electrons
$a=1/\sqrt{n}$.  However, because E-LAT is pinned by impurities, it
cannot adjust its period continuously.  Instead, E-LAT accumulates
stress until it overcomes the pinning force and then makes a sudden
local rearrangement of the lattice, which results in a jump of the
local potential.  Both the E-LOC and E-LAT scenarios are plausible,
and it is difficult to decide between them on the basis of the known
experimental data.  Here we propose a modification of the experiments
\cite{Yacobi00,Yacobi01}, which may help to distinguish between the
two scenarios.

In Ref.\ \cite{Pudalov93}, Pudalov \emph{et al.} observed a very
nonlinear current-voltage ($I$-$V$) relation in the insulating phase
of the 2DEG in Si-MOSFET.  Almost no current $I$ flows until electric
field reaches the threshold field ${\cal E}_t$, and then $I$ sharply
surges at ${\cal E}>{\cal E}_t$, accompanied by the broad-band noise.
Pudalov \emph{et al.} interpreted their findings in terms of
collective sliding of E-LAT depinned by the strong electric field
${\cal E}>{\cal E}_t$, which produces the large current $I$ and
generates the broad-band noise due to the local slip-stick motion.
The $I$-$V$ nonlinearity was found to be extremely sharp, with the
differential conductivity increasing by the factor of $10^6$, in the
samples with the highest mobility and rounded in the samples with poor
mobility \cite{Pudalov-ECRYS}.  These results suggest that the
transition to the insulating state is not driven by disorder, as
assumed by the E-LOC theories, but by E-LAT formation.  The $I$-$V$
nonlinearity was also observed in GaAs samples \cite{Shayegan99}.  It
was shown that the MIT deduced from the temperature dependence of
resistivity is the same one as deduced from the $I$-$V$ nonlinearity
\cite{ShashkinIV}.

We propose to combine the SET experiment with the nonlinear $I$-$V$
experiments.  Suppose a strong pulling electric field ${\cal E}>{\cal
E}_t$ is applied, and E-LAT slides.  Then the SET would register a
time-periodic ac signal with the frequency $\nu=v/l$ produced by E-LAT
of the spatial period $l$, which slides with the velocity $v$.  This
effect is nothing but the narrow-band noise (NBN), well-known for CDW
in the quasi-one-dimensional (Q1D) conductors \cite{Gruner}.  Unlike
in the Q1D conductors, attempts to observe the NBN in regular
transport measurements in the 2DEG failed thus far \cite{Eisenstein}.
We propose that the SET is a better tool for detecting the NBN,
because of its very high sensitivity and because it is a local,
microscopic probe, unlike the macroscopic current leads.  In the
experiment \cite{Yacobi01}, the SET was situated at the distance
$d=400$\,nm from the 2DEG.  This distance is comparable to the average
distance between the carriers $a=1/\sqrt{n}=100$\,nm in the experiment
\cite{Yacobi01} performed on $p$-GaAs with the typical hole
concentration in the insulating phase $n=10^{10}$\,cm$^{-2}$.  Because
$d$ and $l\sim a$ are comparable, the SET should experience a
noticeable time-dependent signal when the periodically-modulated
electron charge density slides past the SET.  Reducing $d$ and
bringing the SET closer to the 2DEG would further increase
sensitivity.

Let us estimate the frequency $\nu=v/l$ of the ac signal.  The E-LAT
period $l$ is of the order of the average distance between the
carriers $l\sim a=1/\sqrt{n}$.  The sliding velocity $v$ is related to
the current density $j=I/W=env$, where $I$ is the total current, and
$W$ is the transverse width of the sample.  Thus we find
\begin{equation}
  \nu \simeq \frac{I}{e}\,\frac{1}{\sqrt{n}W} \simeq 
  \left[6\,\frac{\rm MHz}{\rm pA}\right]
  \frac{I}{\sqrt{n}W}.
\label{f}
\end{equation}
For a crude estimate of the current density in the sliding regime, we
use the data from Ref.\ \cite{Shayegan99} $j=I/W\simeq 0.4\,{\rm
  nA}/0.4\,{\rm mm}=1\,{\rm nA}/{\rm mm}$.  (The data from Ref.\
\cite{Pudalov93} give a similar estimate.)  Substituting these numbers
into Eq.\ (\ref{f}), we find $\nu\simeq 600$\,kHz.  The frequency scale
is similar to that of the Q1D CDW \cite{Gruner}.  Unfortunately, the
frequency range of a typical SET is limited to less than 1\,kHz.
Thus, it is necessary to use the radio-frequency SET (RF-SET)
\cite{Schoelkopf}, which can operate from dc to 100\,MHz.  With this
experimental setup, it should be possible to detect the ac signal at
the frequency $\nu$.

Eq.\ (\ref{f}) shows that $\nu$ is proportional to the current $I$
carried by the sliding E-LAT, and the slope of that dependence is
proportional to $1/\sqrt{n}$.  An experimental observation of this
effect would be the definitive proof of the existence of E-LAT in the
dilute 2DEG.  Periodicity in time is the direct consequence of
periodicity in space, and the E-LOC scenarios cannot produce a
periodic ac signal from the dc current.  Although disorder destroys
the long-range order of E-LAT \cite{Chakravarty,Spivak}, the local
periodicity is preserved and would produce the NBN peak in the Fourier
spectrum.  On the other hand, even if the RF-SET will not find a
time-periodic signal, the measured time series would provide important
microscopic information about electron conduction, such as the
variable-range hopping.  For example, uncorrelated single-electron
hops would generate the Poisson stochastic process in the simplest
case.

\paragraph{Probing spin order with the optically pumped NMR.}

Besides the question of charge ordering in the insulating state of the
MIT, there is a question of spin ordering in that state.  One of the
great tools for obtaining information about electron spins is the
nuclear magnetic resonance (NMR).  In the quantum Hall regime, the
optically-pumped NMR measurements on the $^{71}$Ga nuclei in $n$-GaAs
detected formation of skyrmions in the electron spin configuration for
small deviations from the filling factor $\nu=1$ \cite{Barrett}.  In
the $\nu=1$ state, electrons are spontaneously spin-polarized and
produce a significant effective magnetic field on the nuclei via the
hyperfine interaction.  Thus, the NMR line of the nuclei in contact
with the 2DEG experiences the measurable Knight shift proportional to
the spontaneous spin polarization of electrons
\cite{Barrett,Khandelwal}.

We propose to use a suitable modification of the same method to study
the spin properties of the 2DEG in the insulating state in zero
effective magnetic field.  A magnetic field is needed for NMR, but we
want to eliminate its effect on electrons.  This can be achieved by
engineering a situation where the electron $g$-factor is zero.  For
example, this is the case for a magnetic field parallel to the [100]
surface of $p$-GaAs \cite{Rahimi,g-factor}.  It is also possible to
achieve $g=0$ by applying hydrostatic pressure \cite{W.Kang}.

For the Wigner crystal, different types of spin ordering were proposed
theoretically: ferromagnetic \cite{Zhu}, antiferromagnetic \cite{Zhu},
and various exotic orderings \cite{Chakravarty}.  In the ferromagnetic
state, the NMR line should experience a measurable Knight shift,
detection of which would be the proof of spontaneous spin polarization
of electrons.  In the antiferromagnetic state, the NMR line would
broaden, because the nuclei experience a staggered hyperfine field
from the electrons.  This method is routinely used to detect formation
of spin-density waves (SDW) in Q1D conductors \cite{Gruner}.  On the
other hand, when a strong electric field ${\cal E}>{\cal E}_t$ is
applied, it forces SDW or E-LAT to slide.  Then the nuclei experience
the time-averaged hyperfine magnetic field produced by electrons, and
the NMR line becomes narrow again \cite{Barthel} (the so-called motion
narrowing).  An observation of these effects in NMR would provide a
great deal of information about spin ordering of electrons in the
insulating state and would put the ongoing theoretical discussion of
the subject on a firm experimental ground.

\paragraph{Enhancement of the effective mass and spin susceptibility.}

Experiments
\cite{Kravchenko04,Okamoto,Shashkin01,Vitkalov,Pudalov02,Reznikov}
consistently show that the electron effective mass $m^*$ and the
effective spin susceptibility $\chi^*$ strongly increase when $n\to
n_c$ from the metallic side, where $n_c$ is the critical density of
the MIT.  This phenomenon has a simple explanation within the E-LAT
scenario.  The theory was developed in Refs.\
\cite{Shaginyan,Galitski,Khodel}, and here we only briefly summarize
the main physical idea.

The experiments \cite{Pudalov93,Pudalov-ECRYS} show that the threshold
field ${\cal E}_t$ and the thermal activation gap of resistivity
continuously vanish at $n\to n_c$.  Thus, the phase transition between
the metallic phase at $n>n_c$ and the insulating phase at  $n<n_c$ is
of the second order.  More precisely, it was found to be slightly of
the first order \cite{Pudalov94}, as expected by the symmetry reasons
for a triangular or hexagonal lattice \cite{Brazovskii}.  These
results are in qualitative agreement with the self-consistent
Hartree-Fock calculations \cite{Orozco}, which show that E-LAT
continuously evolves from the CDW limit to the WC limit with the
decrease of $n<n_c$.

Assuming that the system has a tendency to form E-LAT with the wave
vector $q_c\sim1/a$, we can write the charge response function
$S_0(q)=S(q,\omega=0)$ in the following form \cite{Brazovskii} in the
vicinity of the phase transition for $n>n_c$:
\begin{equation}
   S_0(q)\simeq\frac{C_1}{n-n_c + (q-q_c)^2},
\label{eq:S}
\end{equation} 
where $C_1$ is a constant, and $q$ is momentum transfer.  Electrons
can interact via exchange of the critical fluctuations (\ref{eq:S}).
This interaction manifests itself in the Landau interaction function
$f(\theta)\propto S_0(|{\bm p}_1-{\bm p}_2|)$, where ${\bm p}_1$ and
${\bm p}_2$ are the momenta of the interacting electrons, and $\theta$
is the angle between ${\bm p}_1$ and ${\bm p}_2$.  Substituting this
formula in the Landau equation for the effective mass $m^*$ \cite{v9},
we find
\begin{equation}
   \frac{1}{m^*}=\frac{1}{m}
   -C_2\int\frac{\cos\theta\,d\theta}
   {n-n_c + (|{\bm p}_1-{\bm p}_2|-q_c)^2},
\label{eq:m*}
\end{equation}
where $C_2\propto C_1$ is another constant.  Taking into account that
$|{\bm p}_1|=|{\bm p}_2|=p_F$, where $p_F$ is the Fermi momentum, we
obtain
\begin{equation}
   \frac{1}{m^*}=\frac{1}{m}
   -C_2\int\frac{\cos\theta\,d\theta}
   {n-n_c + [2p_F\sin(\theta/2)-q_c]^2}.
\label{eq:m'}
\end{equation}
Assuming that $q_c\sim 2p_F$, we see that the integral in Eq.\
(\ref{eq:m'}) is peaked around $\theta\sim\pi$, where $\cos\theta<0$.
Because of the Fermi statistics, the exchange interaction originating
from the positive Coulomb repulsion is negative, so $C_2<0$.  Thus,
the interaction term in Eq.\ (\ref{eq:m'}) causes an increase in the
effective mass $m^*$.

Moreover, the integral in Eq.\ (\ref{eq:m'}) diverges at $n\to n_c$.
In the case $q_c<2p_F$, it diverges as $(n-n_c)^{-1/2}$
\cite{Shaginyan}:
\begin{equation}
   \frac{m}{m^*}=1-\frac{C_3}{\sqrt{n-n_c}},
\label{eq:m''}
\end{equation}
where $C_3>0$ is another constant.\footnote{In the case $q_c=2p_F$,
the last term in Eq.\ (\ref{eq:m''}) diverges as $(n-n_c)^{-3/4}$.}
As $n$ approaches to $n_c$, the effective mass $m^*$ diverges in Eq.\
(\ref{eq:m''}) even earlier, at $n\to n_\infty=n_c+C_3^2$, where the
last term in Eq.\ (\ref{eq:m''}) becomes equal to 1.  Remember that we
defined $n_c$ in Eq.\ (\ref{eq:S}) as the electron density where E-LAT
forms.  The spin susceptibility is also enhanced via the standard
relation $\chi^*=g^*m^*$.

\begin{figure}[t]
\includegraphics[width=0.8\linewidth,angle=-90]{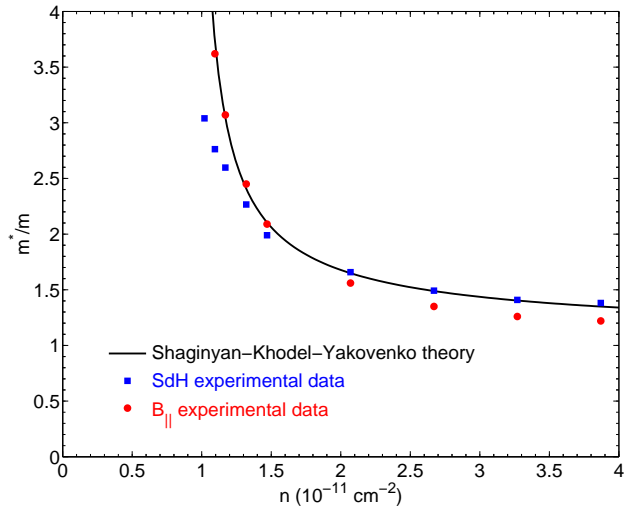} 
\caption{Fig.~1.  Ratio of the effective and bare masses, $m^*/m$,
vs.\ carrier concentration $n$.  The solid line represents the
theoretical formula (\ref{eq:m''}), which is the same as Eq.\ (8) in
Ref.\ \cite{Shaginyan}.  The points represent the experimental data
for Si-MOSFET from Fig.\ 27 of Ref.\ \cite{Kravchenko04}, obtained
from the Shubnikov-de Haas oscillation (SdH, squares) and from the
spin-polarizing parallel magnetic field ($B_\|$, circles).}
\label{fig:m*} 
\end{figure}

\begin{figure}[b]
\includegraphics[width=0.8\linewidth,angle=-90]{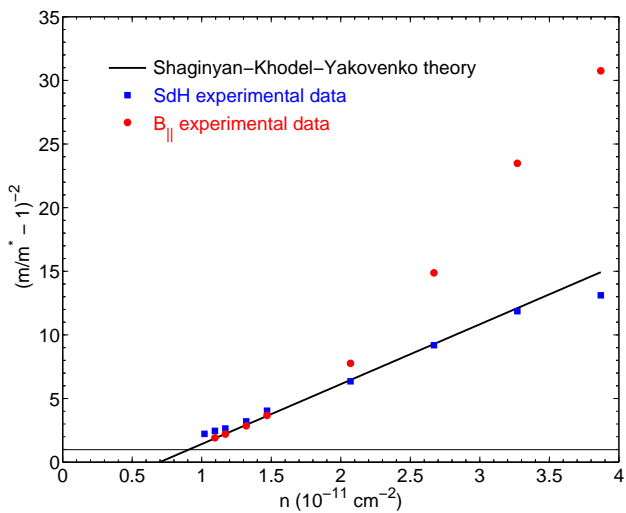} 
\caption{Fig.~2.  $(m/m^*-1)^{-2}$ vs.\ $n$.  The parameters in Eq.\
(\ref{eq:m''}) are determined by the slope and the offset of the
linear fit: $n_c=0.7\times10^{-11}$~cm$^{-2}$ and $C_3=0.46$.  The
effective mass $m^*$ diverges at
$n_\infty=0.9\times10^{-11}$~cm$^{-2}$, where the straight line
crosses level 1 shown by the horizontal line.}
\label{fig:linear} 
\end{figure}

In Fig.\ \ref{fig:m*}, we compare $m^*(n)$ given by Eq.\
(\ref{eq:m''}) with the experimental data for Si-MOSFET from Fig.\ 27
of Ref.\ \cite{Kravchenko04}.  The experimental points were obtained
from the Shubnikov-de Haas oscillation (SdH, squares) and from the
spin-polarizing parallel magnetic field ($B_\|$, circles).  In order
to determine the parameters in Eq.\ (\ref{eq:m''}), we plot
$(1-m/m^*)^{-2}$ vs.\ $n$ in Fig.\ \ref{fig:linear}.  According to
Eq.\ (\ref{eq:m''}), the slope of this linear dependence is $1/C_3^2$,
and it crosses zero at $n=n_c$.  The effective mass $m^*$ diverges at
$n=n_\infty$, where the straight line crosses level 1.  The parameters
of our fit are $n_c=0.7\times10^{-11}$~cm$^{-2}$,
$n_\infty=0.9\times10^{-11}$~cm$^{-2}$, and $C_3=0.46$.

The purpose of Figs.\ \ref{fig:m*} and \ref{fig:linear} is not to
produce a detailed quantitative fit of the experimental data, but only
to demonstrate qualitative agreement between the theory and
experiment.  One should keep in mind that Eq.\ (\ref{eq:m''}) is not
applicable for $n\gg n_c$ far away from the E-LAT transition.  Very
close to the transition point, the singularities predicted by Eq.\
(\ref{eq:m''}) may be cut off by the weakly first-order character of
the phase transition \cite{Pudalov94}.  The divergence of $m^*$ in
Figs.\ \ref{fig:m*} and \ref{fig:linear} is obtained only as an
extrapolation of the available experimental data.  Taking into account
frequency dependence of $S(q,\omega)$ can modify the theory to make
$n_\infty=n_c$.  Nevertheless, the qualitative agreement between the
theory and experiment gives an argument in favor of the E-LAT scenario
for the MIT in the 2DEG.

This theory is also applicable to other system experiencing transition
for a liquid to a crystalline phase.  Such a transition is observed in
the 2D He-3, and the experiment \cite{Casey} finds a very strong
enhancement of $m^*$ in the vicinity of the transition.  Notice that
there is no disorder in liquid He-3, so the E-LOC scenario is
irrelevant in this case.

\paragraph{Conclusions.}

We propose to use the radio-frequency single-electron transistor
(RF-SET) \cite{Schoelkopf} as an extremely sensitive probe
\cite{Yacobi00,Yacobi01} to detect the time-periodic ac signal
generated by sliding electron lattice (E-LAT) at ${\cal E}>{\cal E}_t$
in the insulating state of the 2DEG.  An observation of this
narrow-band-noise effect would be the definitive proof of E-LAT
formation in the dilute 2DEG.  We also propose to use the
optically-pumped NMR technique \cite{Barrett} to probe the electron
spin structure of the insulating state, which may have ferromagnetic,
antiferromagnetic, or exotic types of spin ordering.  NMR can be
performed in a magnetic field without disturbing electron spins in a
situation where the electron $g$-factor is engineered to be zero
\cite{Rahimi,W.Kang}.  Within the Landau theory of Fermi liquids, we
show that critical fluctuations of E-LAT near the metal-insulator
transition produce a strong enhancement of the effective mass $m^*$
and spin susceptibility $\chi^*$ \cite{Shaginyan,Galitski,Khodel} in
qualitative agreement with the experiments in the 2DEG
\cite{Okamoto,Shashkin01,Vitkalov,Pudalov02,Reznikov}, as well as in
the 2D He-3 \cite{Casey}.  This is an argument in favor of the E-LAT
scenario.

Although we concentrated on physics of the 2DEG in zero magnetic
field, the same ideas also apply to the Wigner crystal in a non-zero
magnetic field perpendicular to the 2DEG \cite{1/5,Millis}.

\paragraph{Acknowledgments.}

The authors are grateful to S.~Das Sarma for critical reading of the
manuscript and many valuable comments, to V.~M.~Galitski and
A.~V.~Chubukov for discussions, and to S.~V.~Kravchenko for sending
the experimental data from Fig.\ 27 of Ref.\ \cite{Kravchenko04}.  The
authors thank the Kavli Institute for Theoretical Physics at Santa
Barbara for the opportunity to start this collaboration.  VAK thanks
the Condensed Matter Theory Center for arranging his visit to the
University of Maryland.  VMY is supported by the NSF Grant
DMR-0137726, and VAK by the NSF Grant PHY-0140316, by the McDonnell
Center for Space Sciences, and by the Grant NS-1885.2003.2 from the
Russian Ministry of Industry and Science.

\end{document}